\documentclass{article}
\usepackage{spconf,amsmath,graphicx}
\usepackage[T1]{fontenc}
\usepackage{amssymb}
\usepackage{amsmath}
\usepackage{amsmath,bm}
\usepackage{amsthm}
\usepackage{caption}
\usepackage{graphicx}
\usepackage{subfigure}
\usepackage{cite}
\usepackage{enumerate}
\usepackage[ruled]{algorithm2e}
\usepackage{color, soul}
\usepackage{epstopdf}
\usepackage{url}

\title{Real-Time Prediction for Fine-Grained Air Quality Monitoring System with Asynchronous Sensing}
\name{Zixuan Bai, Zhiwen Hu, Kaigui Bian and Lingyang Song}
\address{School of Electronics Engineering and Computer Science, Peking University, Beijing, China}
\begin{document}
\newcommand{\titlefontsize}{\fontsize{10pt}{10}\selectfont}
%
\maketitle
\begin{abstract}
\vspace{-2mm}
Due to the significant air pollution problem, monitoring and prediction for air quality have become increasingly necessary.
To provide real-time fine-grained air quality monitoring and prediction in urban areas, we have established our own Internet-of-Things-based sensing system in Peking University.
Due to the energy constraint of the sensors, it is preferred that the sensors wake up alternatively in an asynchronous pattern, which leads to a sparse sensing dataset.
In this paper, we propose a novel approach to predict the real-time fine-grained air quality based on asynchronous sensing.
The sparse dataset and the spatial-temporal-meteorological relations are modeled into the correlation graph, in which way the prediction procedures are carefully designed.
The advantage of the proposed solution over existing ones is evaluated over the dataset collected by our air quality monitoring system.
\end{abstract}
\vspace{-2mm}

\vspace{-3mm}
\section{Introduction}\label{sec_Introduction}
\vspace{-2mm}
A recent report from the World Health Organization shows that one in eight of total global deaths are due to air pollution exposure~\cite{WHO}.
Government agencies have defined the air quality index~(AQI) to evaluate pollution degree where
high concentration of fine particles is usually the main factor of a high AQI.
Traditional AQI observation stations can only provide a coarse-grained and high-latency monitoring~\cite{latency}.
However, a recent study shows that the distribution of fine particles could vary within meters~\cite{meters}.

To make up for the above deficiency, it is recommended to deploy numerous low-cost tiny Internet-of-Things~(IoT) sensing devices to monitor fine-grained air quality(mainly $\rm{PM_{2.5}}$ values) for the regions with complicated terrain~\cite{recommended}.
Therefore, we have designed a wireless sensor network system.
The massive commercial IoT sensors can monitor AQIs until the batteries are dead.
In this way, the real-time fine-grained monitoring and prediction are achieved by the massive collected data and the techniques like machine learning\cite{ANNpredict,LSTMpredict}.

However, since the outdoor battery-powered sensors have limited energy\cite{iot1,iot2}, it is preferred that the sensors wake up alternately in an asynchronous way to prolong the lifetime of sensing network~\cite{Hu_jrnl}.
Based on the collected asynchronous data, few existing methods can provide reliable fine-grained AQI prediction.
For instance,
Long Short Term Memory network~(LSTM)~\cite{LSTMpredict} can only be implemented with a synchronous dataset rather than an asynchronous one.
In addition, Multi-layer Perception~(MLP)~\cite{ANNpredict} and spatial-temporal distance weighting interpolation~(IDW)~\cite{IDW} do not perform well based on asynchronous AQI data.
This is because, MLP is unable to reveal the spatial-temporal-meteorological relations among the AQIs in different points-of-interest~(POIs),
and IDW is only an intuitive solution without learning from massive collected data and the influence of the weather conditions.

In this paper, we propose a novel scheme to conduct the fine-grained and real-time prediction of AQI based on asynchronous data collected by our monitoring system.
By designing the correlation graph~(CG), we present the asynchronous sensing data and the spatial-temporal-meteorological relations.
Based on the CG model, the prediction procedures are carefully designed and an optimization problem arises. To obtain the real-time fine-grained prediction results, we aim to solve the optimization problem by an algorithm combining a closed-form derivation and genetic algorithm.
The advantage of the proposed solution over existing ones is evaluated over the dataset collected by our monitoring system.

The main contributions of our work are listed as follows:

1) We present our air quality system in Peking University based on massive IoT sensors, where asynchronous sensing data are modeled into a spatial-temporal-meteorological graph.

2) We propose a novel prediction algorithm for real-time and fine-grained air quality based on sparse dataset.

3) We evaluate the performance gain of our approach over existing methods based on real measured data.

The rest of our paper is organized as follows.
Section~\ref{sec_system_model} introduces the CG model.
Section~\ref{sec_est_steps_and_pro_for} specifies the prediction procedures and applies appropriate methods to obtain real-time and fine-grained prediction.
Section~\ref{sec_Evaluation} shows the simulation results.
Finally, conclusions are given in Section~\ref{sec_Conclusion}.

\vspace{-3mm}
\section{system model}\label{sec_system_model}
\vspace{-2mm}
In this section, we first introduce our air quality asynchronous sensing system in Section~\ref{sec_asyn}. Then, the correlation graph is constructed to model the asynchronous sensing data in Section~\ref{sec_formu}. The spatial-temporal-meteorological relations are added into the CG as weighted edges in Section ~\ref{sec_addedges}.

\vspace{-3mm}
\subsection{Asynchronous Sensing System}\label{sec_asyn}
\vspace{-1mm}
We establish a wireless sensor network system for fine-grained air quality monitoring.
For power efficiency, the deployed IoT sensors are programmed to asynchronously monitor AQI~(mainly $\rm{PM_{2.5}}$ values) until the batteries are dead.

The IoT sensors have been deployed in the campus of Peking University since Feb. 2018. Each sensor uploads the AQI, the location and the time after monitoring~\cite{github}. In addition, weather conditions are collected from the website of China Meteorological Administration~\cite{official}. All features are preprocessed before being applied.
\vspace{-3mm}

\subsection{Correlation Graph for Asynchronous Data}\label{sec_formu}
\vspace{0mm}

\label{model}
\begin{figure}[!thp]
        \vspace{-10mm}
        \centering
        \includegraphics[width=3.4in]{{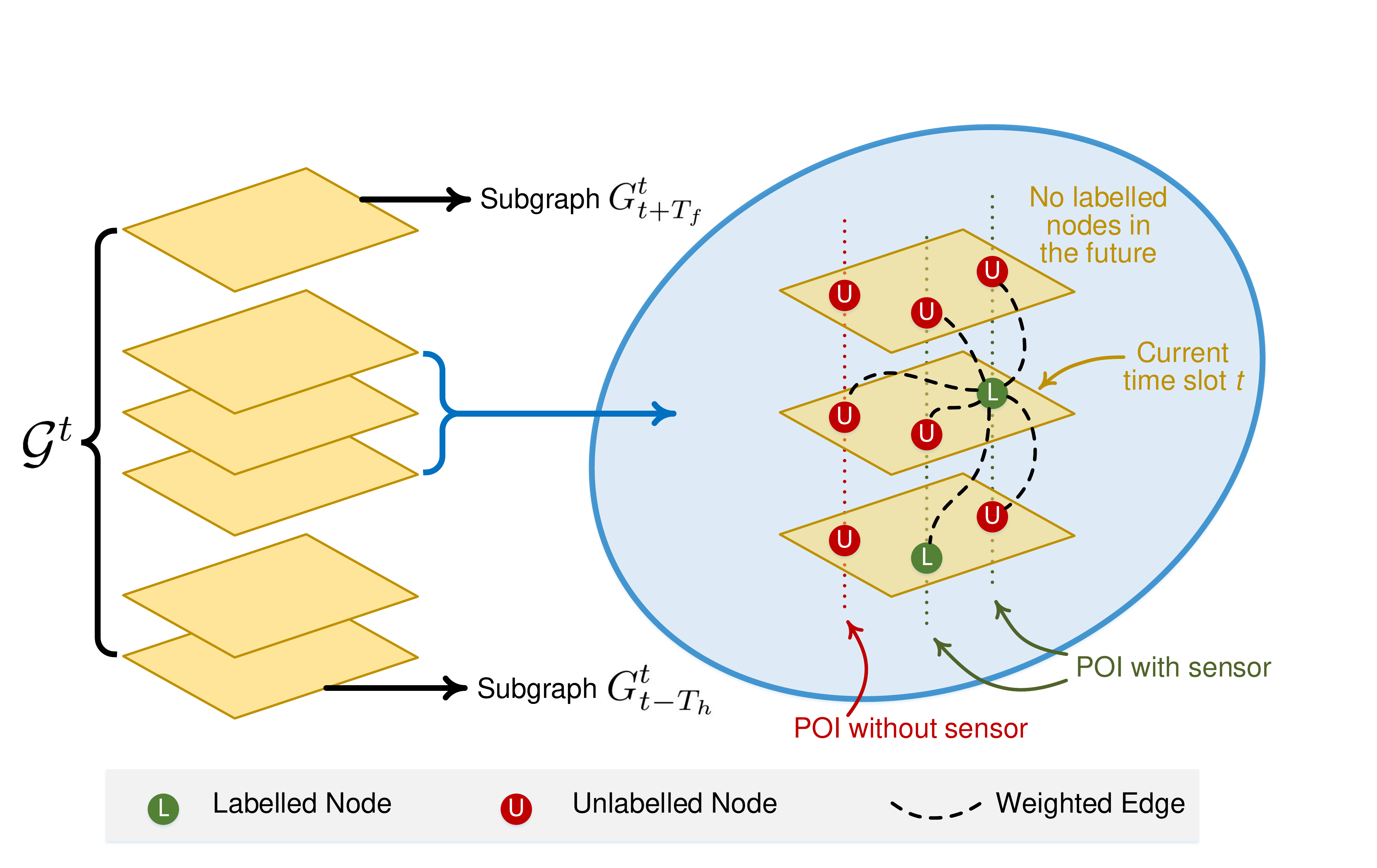}}
        \vspace{-6mm}
        \captionsetup{font={scriptsize}}
        \caption{An illustration of the Correlation Graph.}
\end{figure}
\vspace{-3mm}

There are totally a number of $M$ air quality sensors, denoted as $\{S_1,\cdots,S_m,\cdots,S_M\}$.
Each sensor has a fixed 3D spatial coordinate $(x_m,y_m,z_m)$.
For fine-grained consideration, there are $L$ POIs in concerned areas and we need to know the AQIs of the POIs, where $M < L$.
Due to the limited number of the available sensors, the AQIs of the POIs without the sensors need to be estimated.

Since the sensors monitor data asynchronously, we divide the system into short equal-length time slots, where each data can be considered as collected in a specific time slot.
The collected data therefore could be very sparse, because only a small proportion of the sensors collect and upload data at any certain time slot.

At time $t$, we construct correlation graph $\mathcal{G}^t$ to illustrate the state of the system.
$\mathcal{G}^t$ has multiple subgraphs, denoted as $\{G^t_{t^{\prime}}\}$.
Each subgraph $G^t_{t^{\prime}}$ contains $L$ nodes, where each node implies the collected value or the estimated value of a specific POI at time $t^{\prime}$.
The node is \textit{labeled} if the corresponding data is collected by the sensor, otherwise, the node is \textit{unlabeled}.
The sets of the \textit{labeled} and \textit{unlabeled} nodes are given by $\mathcal{L}^t$ and $\mathcal{U}^t$ respectively.
Each \textit{labeled} node $v^t_n$ is combined with a measured value denoted as $c^t_n$.
For each $\mathcal{G}^t$, we only consider limited number of subgraphs, given by $\{G_{t-T_h}^t,\cdots,G_{t+T_f}^t\}$, where $T_h$ is the number of concerned historical subgraphs from current time $t$ and $T_f$ is the number of concerned future subgraphs from current time $t$.
Therefore, the total number of the nodes in $\mathcal{G}^t$ is $N = (T_h+T_f+1) \times L$.
The set of all the nodes is denoted as $\mathcal{V}^t=\{v_1^t,\cdots,v_n^t\cdots,v_N^t\}= \mathcal{L}^t \bigcup \mathcal{U}^t$.
We define the real-time fine-grained prediction as $\mathcal{F}^t=[F_1^t,\cdots,$
$F_n^t,\ldots,F_N^t]^{\rm{T}}$, where $F_n^t$ denotes the inferred value for node $v_n^t$ at time $t$.

\vspace{-3mm}
\subsection{Feature Relations in Correlation Graph}\label{sec_addedges}
\vspace{-1mm}
Any of two nodes are connected by an edge. The weight of the edge depends on the spatial-temporal relation of the two nodes, as well as the weather conditions of the corresponding time slots.

\textbf{The weighted feature vector:}
We first define the spatial-temporal-meteorological feature vector of node $v_n^t$ as
\vspace{-3mm}
\begin{equation}
\vspace{-2mm}
\bm{q}_{n}^t = \left[q_{n,1}^t,\cdots,q_{n,k}^t,\cdots,q_{n,K}^t\right]^{\rm T} \label{feature_vector},
\end{equation}
where $K$ is the number of features and $q_{n,k}^t$ is the \textit{k}-th feature of $v_n^t$.
The feature vector $\bm{q}_n^t$ contains the spatial coordinates, the temporal coordinate of $v_n^t$, and the meteorological conditions, including weather types, wind speeds, wind directions, temperature and
humidity at time slot $t$.
Due to the fact that different elements in $\bm{q}_n^t$ may have different impacts on the relations between two nodes, we introduce a weight vector $\bm{\beta}\!=\!{\left[\beta_1,\cdots,\beta_{k},\cdots,\beta_K \right]}^{\rm T}$, where $\beta_k \in (0,1),\forall k \!\in\! \left\{ 1,2,\cdots,K \right\}$ and ${||\bm{\beta}||}_1\!=\!\sum\nolimits_{i\!=\!1}^K{|\beta_k|} \!=\! \sum\nolimits_{i\!=\!1}^K{\beta_k} \!=\! 1$.
Then the weighted feature vector $\bm{f}_n^t \!=\! [f_{n,1}^t,\cdots,f_{n,k}^t,\cdots,f_{n,K}^t]^{\rm T}$ of node $v_n^t$ is expressed as
\vspace{-4mm}
\begin{equation}
\vspace{-2mm}
{\bm{f}_n^t}=\bm{q}_n^t \odot \bm{\beta}=\left[q_{n,1}^t\beta_{1},\cdots,q_{n,k}^t\beta_{k},\cdots,q_{n,K}^t\beta_{K}\right]^{\rm T} \label{weighted_feature_vector},
\end{equation}
where $\odot$ denotes hadamard product.

\textbf{Similarity of two nodes:}
We define the adjusted cosine similarity $m_{ij}^t$ between $v^t_i$, $v^t_j$, which is given by
\vspace{-3mm}
\begin{equation}
\vspace{-2mm}
m_{ij}^t = \frac{\langle \bm{f}_i^t-\overline{\bm{f}},\bm{f}_j^t-{\overline{\bm{f}}} \rangle}{\sqrt{\left\|\bm{f}_i^t- \overline{\bm{f}}\right\|_2^2}{\sqrt{\left\|\bm{f}_j^t-\overline{\bm{f}}\right\|_2^2}}},
\end{equation}
where $\overline{\bm{f}}$ denotes the mean values of the weighted features.

\textbf{The weight on each edge:}
For each node, its \textit{k}-nearest neighbors are defined as the nodes that have the top \textit{k} highest similarity with this certain node.
If node $v_j^t$ is one of $v_i^t$'s \textit{k}-nearest neighbors \textbf{or} $v_i^t$ is one of $v_j^t$'s \textit{k}-nearest neighbors, we define the weight of the edge between $v_i^t$, $v_j^t$ as
\vspace{-3mm}
\begin{equation}
\vspace{-2mm}
w_{ij}^t = \dfrac{1}{2} {\tanh\Big(\alpha_1(m_{ij}^t-\alpha_2)\Big)+\dfrac{1}{2}},
\end{equation}
where the parameter $\alpha_1(\alpha_1>0)$ magnifies the differences of similarities and the parameter $\alpha_2 \in \left(-1,1\right)$ is a cutoff value.
$w_{ij}^t$ is in $\left(0,1\right)$ since it is a monotone function and $m_{ij}^t$ ranges from $-1$ to $1$.
If node $v_i^t$, $v_j^t$ are not one of the \textit{k}-nearest neighbors of each other, the weight $w_{ij}^t$ between them is set to 0 to reduce the influence by dissimilar nodes and to improve computing efficiency.
It is worth stressing that $v_i^t$ is not a neighbor of itself and we simply set $w_{ii}^t=0, \forall v_i^t \in \mathcal{V}^t$.

Further, a real symmetric weight matrix $W^t$ can then be given by $W^t=\left[W^t\right]^{\rm{T}}=\left[w_{ij}^t\right]_{N \times N}$. In addition, the degree $d_{n}^t$ of node $v_n^t$ is given by $d_{n}^t=\sum\nolimits_{i=1}^N{w_{in}^t}=\sum\nolimits_{i=1}^N{w_{ni}^t}$.

\section{Prediction Procedures Design}\label{sec_est_steps_and_pro_for}
\begin{figure}[!thp]
    \centering
    \includegraphics[width=3.5in]{{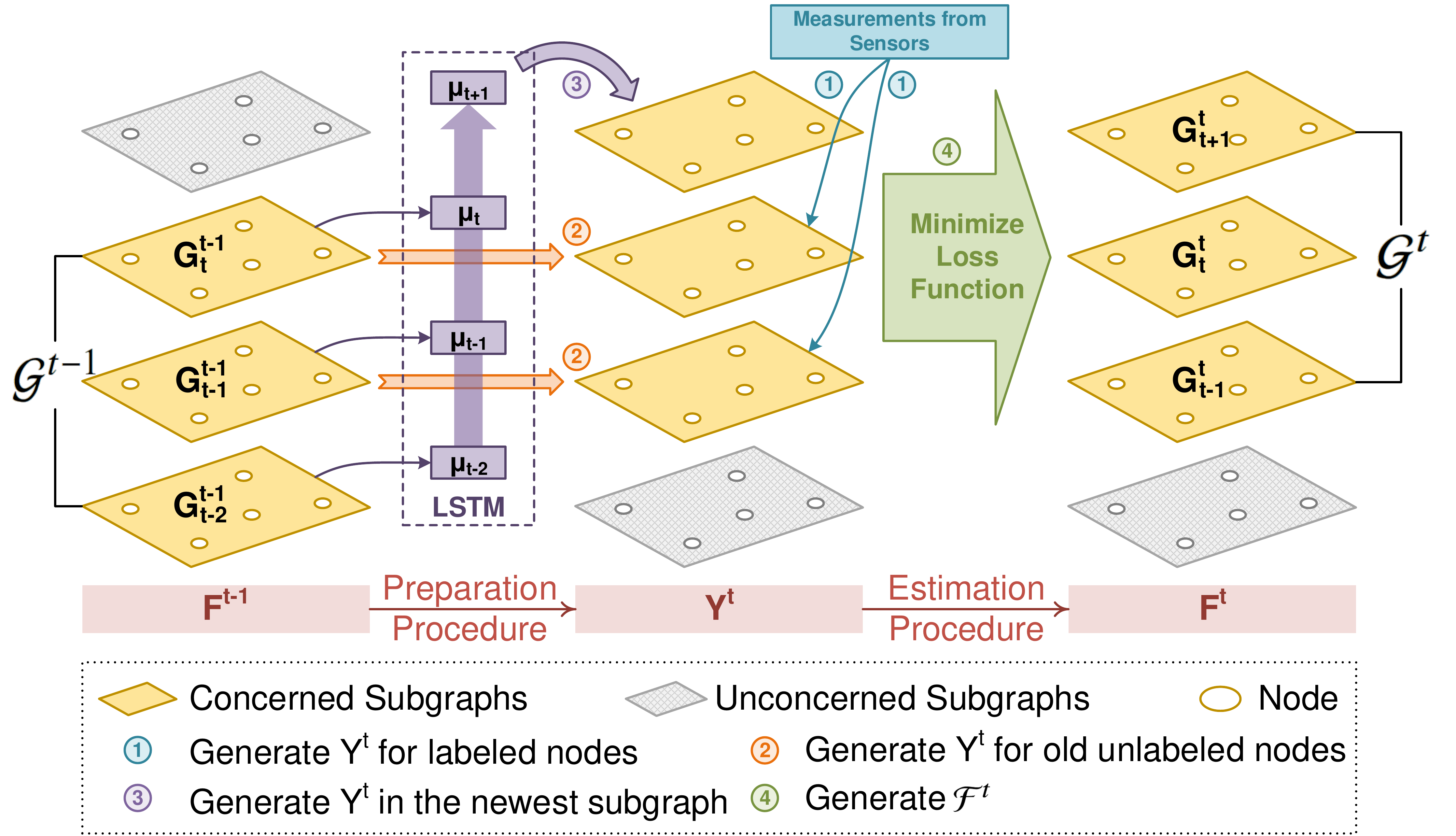}}
    \vspace{-6mm}
    \captionsetup{font={scriptsize}}
    \caption{An illustration of the prediction procedure, where \textcircled{1}, \textcircled{2} and \textcircled{3} show the preparation procedure to obtain the pre-estimation vector $\mathcal{Y}^t$, and \textcircled{4} shows the estimation procedure to obtain the final estimation result $\mathcal{F}^t$.}\label{update}
    \vspace{-4mm}	
\end{figure}

\vspace{-3mm}
After receiving the monitoring data at time slot $t$, the prediction system immediately executes a round of iteration in order to obtain the real-time prediction $\mathcal{F}^t$, by using the history prediction $\mathcal{F}^{t-1}$ and the measured data $\{c^t_n\}$ for the \textit{labeled} nodes.
This includes two procedures, namely the preparation procedure and the estimation procedure, as shown in Fig.~\ref{update}.

\vspace{-3mm}

\vspace{-5mm}
\subsection{Preparation Procedure}
\vspace{-2mm}
Based on $\mathcal{G}^{t-1}$ and $\mathcal{F}^{t-1}$ for the last time slot $t\!-\!1$, and the measured values $\{c^t_n\}$ at the current time slot $t$, we first construct a pre-estimation vector for the current time, given by $\mathcal{Y}^t=[Y_1^t,\cdots,Y_n^t,\cdots,Y_N^t]^{\rm{T}}$.
$\mathcal{Y}^t$ only shows the rough values of the nodes in $\mathcal{G}^t$ and needs to be refined in the subsequent estimation procedure.

\textbf{For the \textit{labeled} nodes:}
We set $Y_n^{t} = c^t_n$ for node $v^{t}_n$.

\textbf{For the \textit{unlabeled} nodes not in the newly-added subgraph:}
When it comes to the \textit{t}-th time slot, $\mathcal{G}^{t}$ is generated by removing the oldest subgraph and adding a new subgraph for time $t+T_f$.
The \textit{unlabeled} nodes not in the newly-added subgraph have been estimated in the last-time prediction.
Therefore, for the node $v_n^t$ in these subgraphs, we set $Y^{t}_n$ the same to the last-time estimated value $F^{t-1}_{n^{'}}$~(Suppose the \textit{n}-th node in $\mathcal{G}^t$ is the $n^{\prime}$-th node in $\mathcal{G}^{t-1}$).

\textbf{For the \textit{unlabeled} nodes in the newly-added subgraph: }
An LSTM~\cite{LSTM} network is used to determine their pre-estimated values.
Specifically, the mean value $\mu_{\tau}$ for each monitoring time slot $\tau$ is calculated first to represent the coarse-grained estimation, given by
\vspace{-4mm}
\begin{equation}
\vspace{-2mm}
\mu_{\tau} = \overline{F_{\tau}},\quad \tau = {1,2,\cdots,t\!-\!1\!+\!T_f},
\end{equation}
where $\overline{F_{\tau}}$ denotes the average of the last-time estimated values for time slot $\tau$.
With the meteorological features and a series of the coarse-grained estimations as the training dataset, an LSTM network can be trained in advance and can be immediately used to predict the coarse-grained estimation $\mu_{t+T_f}$ in the newly-added subgraph.
Finally, we set $\mu_{t+T_f}$ as the pre-estimations for all the nodes in $G^{t}_{t+T_f}$.

To sum up, the pre-estimation is given by
\vspace{-1mm}
\begin{equation}
\vspace{-2mm}
    Y^{t}_n=
    \left\{
        \begin{array}{ll}
            c^{t}_n,&  v_n^{t} \in \mathcal{L}^{t},\\
            F^{t-1}_{n^{'}} ,& v_n^{t} \in \mathcal{U}^{t} \text{ and } v_n^{t} \not\in G^{t}_{t+T_f},\\
            \mu_{t+T_f},& v_n^{t} \in G^{t}_{t+T_f}.
        \end{array}
        \right.\label{subgraph_coarse-grained_estimation}
\end{equation}
The steps \textcircled{\small{1}}, \textcircled{\small{2}}, and \textcircled{\small{3}} in Fig.~\ref{update} illustrate the three kinds of assignments in~(\ref{subgraph_coarse-grained_estimation}) respectively.

\vspace{-3mm}
\subsection{Estimation Procedure}
\vspace{-2mm}
Based on the pre-estimated values in $\mathcal{Y}^t$, we need to calculate more precise prediction values in $\mathcal{F}^t$.

\vspace{-4mm}
\subsubsection{Minimizing the loss function}
\vspace{-2mm}
Two aspects are concerned when calculating $\mathcal{F}^t$.
The first one is smoothness, implying that two closely related nodes (with a high weight edge between them) need to have similar values.
The second one is reliability, indicating that the final estimation and the pre-estimation should be as same as possible, given by $\mathcal{F}^t \approx \mathcal{Y}^t$.

Combining \textit{smoothness} and \textit{reliability}, we design the loss function with the help of~\cite{zhou2004learning}, given by:
\vspace{-3mm}
\begin{equation}
\vspace{-3mm}
    \mathbb{L}^t(\mathcal{F}^t\!\!,\bm{\beta})\!=\!\frac{1}{2}\!\!\!\!\!\sum\limits_{v_i^t,v_j^t \in \mathcal{V}^t} \!\!\!\!\! w_{ij}^t \bigg( \frac{F_i^t}{\sqrt{d_i^t}} -\! \frac{F_j^t}{\sqrt{d_j^t}}\bigg)^2\!\!\!\!+\!\lambda\!\!\! \sum\limits_{v_i^t \in \mathcal{V}^t}\!\! \big( F_i^t\!-\!Y_i^t \big)^2 \label{loss_function},
\end{equation}
where $\lambda$ is the balanced parameter. The first term on the right side in (\ref{loss_function}) is the \textit{smoothness} term. Instead of using the differences of the function values on the two neighboring nodes directly, we regularize each function value $F_i^t$ with the corresponding normalized coefficient ${1}/{\sqrt{d_i^t}}$. The second term on the right side in (\ref{loss_function}) is the \textit{reliability} term.

It is worthy to note that $\bm{\beta}$ does not need to be updated frequently, since the weights of different features are supposed to be stable in a short time.
Our objective is to minimize the average of the loss functions in $\mathcal{T}$, given by
\vspace{-2mm}
\begin{equation}
\vspace{-2mm}
    \begin{split}
    &\mathop{\arg\min}_{\bm{\beta},\{\mathcal{F}^t\}} \frac{1}{|\mathcal{T}|}\sum\limits_{t \in \mathcal{T}}{\mathbb{L}^t(\bm{\beta},\mathcal{F}^t)},\\
    s.t. \quad&{||\bm{\beta}||}_1=1,\quad\beta_k>0,\quad\forall \beta_k \in \{1,\cdots,K\}, \label{optimization}
    \end{split}
\end{equation}
where $\mathcal{T}$ represents a set of the history monitoring time slots.
Note that the iterative utilization of the estimated \textit{unlabeled} nodes for each next round forms a semi-supervised learning approach.

The minimization problem in $(\ref{optimization})$ can be divided into two optimization sub-problems. Supposing that $\bm{\beta}$ is fixed, the first sub-problem is given by
\vspace{-2mm}
\begin{equation}
\vspace{-2mm}
    \mathbb{L}_1^t(\bm{\beta})=\min_{\mathcal{F}^t} \mathbb{L}^t(\mathcal{F}^t,\bm{\beta}) \label{optimization_1}.
\end{equation}
Correspondingly, the second sub-problem is given by
\vspace{-2mm}
\begin{equation}
\vspace{-2mm}
\begin{split}
    &\mathop{\arg\min}_{\bm{\beta}} \frac{1}{|\mathcal{T}|}\sum\limits_{t \in \mathcal{T}}{\mathbb{L}^t_1(\bm{\beta})},\\
    s.t. \quad&||\bm{\beta}||_1=1,\quad\beta_k>0,\quad\forall \beta_k \in \{1,\cdots,K\}
    \label{optimization 2}.
\end{split}
\end{equation}

\vspace{-4mm}
\subsubsection{Solution to sub-problem~(\ref{optimization_1})}
\vspace{-2mm}
$\mathbb{L}^t(\mathcal{F}^t,\bm{\beta})$ in (\ref{loss_function}) is a convex function when $\bm{\beta}$ is given. We differentiate $\mathbb{L}^t$ with respect to $\mathcal{F}^t$ and set it to $\bm{0}$, given by
\vspace{-3mm}
\begin{equation}
\vspace{-3mm}
    \frac{\partial \mathbb{L}^t}{\partial \mathcal{F}^t}
    =\left[\frac{\partial \mathbb{L}^t}{\partial F_1^t},
    \frac{\partial \mathbb{L}^t}{\partial F_2^t},
    \cdots,
    \frac{\partial \mathbb{L}^t}{\partial F_N^t}
    \right]^{\rm{T}}=\bm{0}^{\rm{T}}.
\end{equation}

After derivation, the final solution of $\mathcal{F}^t$ is\footnote{Since $\left|I-\frac{1}{1+\lambda}{(D^t)}^{-\frac{1}{2}}W^t{(D^t)}^{-\frac{1}{2}}\right|\neq 0$, the matrix is invertible.} given as
\vspace{-3mm}
\begin{equation}
\vspace{-3mm}
    \mathcal{F}^t=\frac{\lambda}{1+\lambda}\left(I-\frac{1}{1+\lambda}{(D^t)}^{-\frac{1}{2}}W^t(D^t)^{-\frac{1}{2}}\right)^{-1}\mathcal{Y}^t \label{solution},
\end{equation}
where $(D^t)^{-\frac{1}{2}}=diag({1}/{\sqrt{d_1^t}},\cdots,{1}/{\sqrt{d_N^t}})$.

It is noteworthy that once $\bm{\beta}$ is figured out, we can directly obtain the real-time fine-grained prediction according to~(\ref{solution}).

\vspace{-4mm}
\subsubsection{Solution to sub-problem~(\ref{optimization 2})}
\vspace{-2mm}
The value of $\bm{\beta}$ is obtained by the following genetic algorithm based on long-term weather conditions and the measured air quality data.

Each individual $\mathcal{P}(n)$ in the generation set $\mathcal{P}$ represents a trial solution to (\ref{optimization 2}) and we encode it to a series of binary digits $\mathcal{P}(n)=(s_1^n,s_2^n,\cdots,s_{K\times R}^n)$ as its genotype, where $R$ denotes the encoding length of each element in $\bm{\beta}$ and $K$ is the number of features. Therefore, $K\times R$ denotes the total length of an individual's genotype.
The genetic algorithm is specified as follows.

\textbf{Crossover:} For any two individuals in the old generation, crossover can be performed with a fixed probability $p_c$ to produce new individuals. To be specific, for individuals $\mathcal{P}(i)$ and $\mathcal{P}(j)$, we randomly choose a position $p\!=\!2,3,\cdots,K\times R$ and then the genotypes for the new individuals are denoted as $\mathcal{P}(i^{'}) \!=\! (s_1^i,\cdots,s_{p-1}^i,s_{p}^j,\cdots,s_{K\times R}^j)$ and $\mathcal{P}(j^{'}) = (s_1^j,\cdots,s_{p-1}^j,s_{p}^i,\cdots,s_{K\!\times\! R}^i)$ after crossover.

\textbf{Mutation:} Each binary digit in the old generation after crossover remains the same in most of time and alters to the opposite with a fixed probability $p_m$.

\textbf{Selection:} The fitness value for each individual in the old generation is given by
\vspace{-3mm}
\begin{equation}
\vspace{-3mm}
    z\big(\mathcal{P}(n)\big) =1\big/\left(
    \frac{1}{|\mathcal{T}|}\sum\limits_{t \in \mathcal{T}}{\mathbb{L}^t_1(\bm{\beta})}\right),\  \forall \mathcal{P}(n) \in \mathcal{P}
    \label{fitness}.
\end{equation}
The new generation is chosen from the individuals after crossover and mutation according to their survival probabilities where the survival probability for each individual is in proportion to its fitness value.

\textbf{Evolution Overview:}
Crossover and mutation are first performed to create new individuals in the old generation.
To satisfy the constraint ${||\bm{\beta}||}_1=1$ for the new individuals, we normalize them and it is easy to prove that the fitness values won't change after normalization.
The new generation is then produced by the selection procedure with constraining the size of the new generation equals to the size of the old one.
The genetic algorithm is terminated when a better individual hasn't appeared in the latest $E$ generations.
\vspace{-3mm}
\section{Evaluation}\label{sec_Evaluation}
\vspace{-2mm}

\subsection{Parameter Setup}
\vspace{-1mm}



In asynchronous sensing, the probability of each sensor detecting data at each time slot is simply set to ${1}/{5}$ to illustrate the most general case.
The other parameters are listed in Table~\ref{tab_Parameters}.

\vspace{3mm}
\begin{table}[!thp]
    \renewcommand\arraystretch{0.9}
    \caption{Parameter Setup}\label{tab_Parameters}   \centering
    \vspace{-2mm}
    \begin{tabular}{|p{54mm}|p{23mm}|}
    \hline
    Number of nodes in one subgraph $L$ & 60\\
    \hline
    Number of sensors in one subgraph $M$ & between 5 to 30\\
    \hline
    Similarity parameters $\alpha_1$ and $\alpha_2$ & $20$ and $0$ \\
    \hline
    Number of history subgraphs $T_h$ & between $3$ to $8$ \\
    \hline
    Number of future subgraphs $T_f$ & between $1$ to $7$ \\
    \hline
    Encoding length $R$ & $20$\\
    \hline
    Parameter $k$ in \textit{k}-NN & $200$ \\
    \hline
    Trade-off parameter $\lambda$ & $0.3$\\
    \hline
    Crossover probability $p_c$ & $0.6$\\
    \hline
    Mutation probability $p_m$ & $0.05$\\
    \hline
    Termination conditions $E$ & $500$\\
    \hline
    Length of each time slot  & $5$ minutes\\
    \hline
    \end{tabular}
    \vspace{-3mm}
\end{table}
\vspace{-3mm}
\subsection{Simulation Results and Discussions}\label{sec_simulations}

\begin{figure}[!thp]
\vspace{-2mm}
  \centering
  \includegraphics[width=3.6in]{{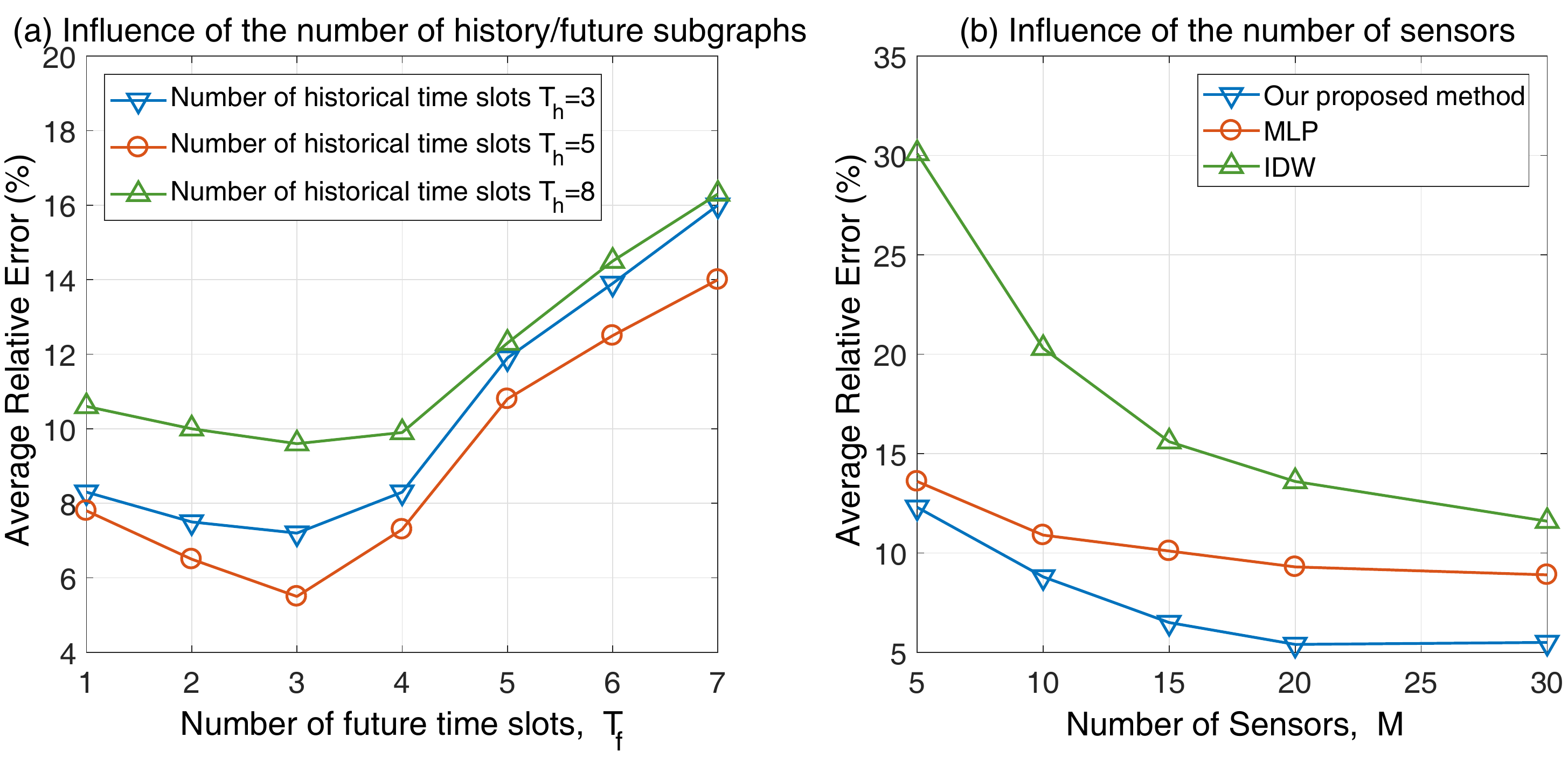}}
  \captionsetup{font={scriptsize}}
  \vspace{-6mm}
  \caption{The performance of the proposed solution with respect to different settings.}\label{simulation_graph}
  \vspace{-2mm}
\end{figure}

In Fig.~\ref{simulation_graph}(a), we set $M\!=\!30$ and compare the performances when the numbers of the history/future subgraphs change, where the performances are indicated by average relative prediction errors.
The results indicate that the appropriate numbers of the history/future subgraphs lead to a better performance.
For the history subgraphs, $\mathcal{G}^t$ may lack history information when $T_h$ is too small, and $\mathcal{G}^t$ will contain worthless information if $T_h$ is too big.
For the future subgraphs, $\mathcal{G}^t$ may have insufficient meteorological conditions in the near future when $T_f$ is too small, and $\mathcal{G}^t$ will contain relatively few \textit{labeled} nodes if $T_f$ is too big.

In Fig.~\ref{simulation_graph}(b), we compare the performances of different approaches, including Multi-layer perception~(MLP), Inverse distance weighting interpolation~\cite{IDW}, and the proposed CG Model (with $T_h\!=\!5$, $T_f\!=\!3$).
A significant performance gain of our solution can be observed, and the average relative error can be reduced to $5\%$ when there are enough sensors.


\vspace{-3mm}
\section{Conclusion}\label{sec_Conclusion}
\vspace{-2mm}
In this paper, we modeled the asynchronous air quality data and their spatial-temporal-meteorological relations by using the weighted CG.
A novel real-time fine-grained prediction method was proposed based on the sparse dataset.
Simulation results show that the size of the CG should not be too big or too small.
The proposed scheme reduces the error to 5$\%$ (with $T_h\!=\!5$, $T_f\!=\!3$, $M\!=\!30$) while the error of MLP is about $9\%$ in asynchronous sensing situation.

\vfill\pagebreak


\end{document}